\documentclass[twocolumn,showpacs,preprintnumbers,amsmath,amssymb]{revtex4-1}
\usepackage{graphicx}
\usepackage{color}
\usepackage[normalem]{ulem}  % \sout{old text} for strikeout

\renewcommand\sout{\bgroup \color{red}\ULdepth=-.5ex \ULset}
\newcommand\soutb{\bgroup \color{blue} \ULdepth=-.5ex \ULset}

\usepackage{graphicx}% Include figure files
\usepackage{dcolumn}% Align table columns on decimal point
\usepackage{bm}% bold math

\begin{document}

\preprint{}
\def\vc#1{\mbox{\boldmath $#1$}}

\title{Electric dipole moment of the deuteron in the standard model with $NN - \Lambda N - \Sigma N$ coupling}% Force line breaks with \\

\author{Nodoka~Yamanaka$^1$}
  \email{nodoka.yamanaka@riken.jp}
  \affiliation{$^1$iTHES Research Group, RIKEN, %\\
  Wako, Saitama 351-0198, Japan}

\date{\today}% It is always \today, today,
             %  but any date may be explicitly specified

\begin{abstract}

We calculate the electric dipole moment (EDM) of the deuteron in the standard model with $|\Delta S| =1$ interactions by taking into account the $NN - \Lambda N - \Sigma N$ channel coupling, which is an important nuclear level systematics.
The two-body problem is solved with the Gaussian Expansion Method using the realistic Argonne $v18$ nuclear force and the $YN$ potential which can reproduce the binding energies of $^3_\Lambda$H, $^3_\Lambda$He, and $^4_\Lambda$He.
The $|\Delta S| =1$ interbaryon potential is modeled by the one-meson exchange process.
It is found that the deuteron EDM is modified by less than 10\%, and the main contribution to this deviation is due to the polarization of the hyperon-nucleon channels.
The effect of the $YN$ interaction is small, and treating $ \Lambda N$ and $ \Sigma N$ channels as free is a good approximation for the EDM of the deuteron.

\end{abstract}

% PACS, the Physics and Astronomy
                             % Classification Scheme.
%CP invariance, 11.30.Er
%Electric moments, nuclear, 21.10.Ky
%Nuclear tests of fundamental interactions and symmetries, 24.80.+y
%Hypernuclei, 21.80.+a

%\keywords{Suggested keywords}%Use showkeys class option if keyword
                              %display desired
\maketitle

%%%%%%%%%% Introduction %%%%%%%%%%%%%%%

\section{Introduction\label{sec:introduction}}

From recent observations, 5\% of the energy of our Universe is composed of baryonic matter \cite{planck}.
To generate the baryon number asymmetry, CP violation is required, according to Sakharov's criteria \cite{sakharov}.
In the standard model (SM) of particle physics, however, the baryon-to-photon ratio generated by the CP phase of the Cabibbo-Kobayashi-Maskawa (CKM) matrix \cite{ckm} is only $1:10^{20}$ \cite{farrar}, being in significant deficit compared with the observed data $1:10^{10}$ \cite{planck}.
We therefore need to extend the SM with a new theory containing additional source(s) of larger CP violation.

A promising experimental approach to search for CP violation is the measurement of the {\it electric dipole moment} (EDM) \cite{edmreview,khriplovichbook,ginges,pospelovreview,hewett}.
The EDM is an observable very sensitive to the CP violation. 
Its first advantage is its accurate measurability in experiments.
The second one is that its measurement costs much less than accelerator experiments.
The third strong point is the variety of the systems for which the EDM can be measured.
Currently, the experimental studies of the EDMs of the neutron \cite{baker}, atoms \cite{rosenberry,regan,griffith,graner,parker,bishof}, molecules \cite{hudson,acme} and muon \cite{muong2} have been performed, and many future projects are also developed.

Among the systems where the EDM can be measured, the nuclear EDM is of particular interest \cite{hewett,storage1,storage2,storage3,storage4,storage5,storage6,storage7,storage8,storage9,storage10}.
Recently, the measurement of the EDM of light nuclei using storage rings is planned, and an impressive sensitivity of $O(10^{-29}) e$ cm is estimated \cite{storage8,storage9,storage10,bnl}.
The nuclear EDM also has advantages of its own.
First, the nuclear EDM does not suffer from Schiff's screening \cite{schiff}, since we have no electrons to screen the charged ions.
It may also enhance the nucleon level CP violating effect due to the many-body effect \cite{sushkov,yamanakanuclearedm}.
Owing to these arguments, the nuclear EDM is a very attractive probe of the CP violation of strong interacting sector, and theoretical investigations \cite{korkin,liu,pospelovdeuteron,afnan,devries,dedmtheta,stetcu,chiral3nucleon,song,bsaisou,bsaisou2,yamanakanuclearedm,mereghetti,13cedm} were extensively done so far.

What about the SM contribution?
In the SM, the CP phase of the CKM matrix contributes to the EDM \cite{ellis,khriplovichnedm,avishai,avishai2,ellisthetasm,khriplovichtheta,gerard,sushkov,donoghue,mckellar1,smeedm,hesmedm,smweinbergop,czarnecki,mannel,seng,yamanakasmedm}.
This is an important background in the search for new physics.
It was long thought that the EDM in the SM is small.
In a recent work, it was shown that the EDM of light nuclei generated by the $|\Delta S| = 1$ long distance effect of the SM may contribute at the level of $O(10^{-31})e$ cm \cite{yamanakasmedm}.
This value is below the prospective experimental sensitivity, but it is not sizably smaller than it.
If the SM contribution is enhanced in nuclei, the EDM may also become a good probe of CKM unitarity.
It is therefore of importance to study its precise value and the systematics.

In previous studies \cite{sushkov,hesmedm,yamanakasmedm}, the P, CP-odd $NN$ interaction was given as a combination of the $\Delta S = +1$ and $\Delta S = -1$ hadron level interactions, respecting the Jarlskog combination \cite{jarlskog}, and the free hyperon of the intermediate state was integrated out.
The dynamical effect of the strange intermediate states was actually neglected.
As the $|\Delta S| = 1$ transitions are weak processes, the strong dynamics of the intermediate nuclear states with hyperons may potentially be important, and there are no firm reasons to neglect it.
As this effect constitutes an important nuclear level systematics in the evaluation of the nuclear EDM, we have to discuss it in detail.
For this purpose, we first investigate the dynamical effect of the strange intermediate state on the deuteron, the simplest nuclear system.
The dynamical effect of the strange intermediate state is described in the hypernuclear physics.
The interaction between nucleons and hyperons ($YN$ interaction) was extensively discussed in the past using hypernuclei
\cite{motoba1,motoba2,bando,hiyama,hiyamahypernuclei1,hiyamaYNLS,Nemura2002,hiyamaYN1,hiyamadoublelambda,hiyamahypernuclei2,hiyamaYN2}.
This problem can be solved using the Gaussian Expansion Method \cite{hiyama}, which was used to describe many systems \cite{3nucleon,benchmark,gemapplication}.

This paper is organized as follows.
In the next section, we first briefly review our calculational method, the Gaussian Expansion Method, and present the interactions adopted in this discussion.
The derivation of the $|\Delta S|=1$ weak interaction is presented in detail, with the tree level couplings borrowed from the previous work. 
We then define in Section \ref{sec:edm} the EDM.
In Section \ref{sec:analysis}, the results of the evaluation of the EDM of the deuteron in the SM are presented, and we analyze them in detail.
We finally summarize our discussion.

%%%%%%%%%% Method and interaction %%%%%%%%%%%%%%%

\section{Method and interaction}

\subsection{Gaussian Expansion Method}

To solve the Schr\"{o}dinger equation of the deuteron, we use the Gaussian Expansion Method \cite{hiyama}.
The Gaussian Expansion Method can solve the nonrelativistic Schr\"{o}dinger equation 
\begin{eqnarray}
( H - E ) \, \Psi_{JM_z,TT_z}  = 0 ,
\label{eq:schr7}
\end{eqnarray}
by diagonalizing the hamiltonian $H$ expressed in the Gaussian basis.
The deuteron wave function is given as
\begin{eqnarray}
&&\Psi_{JM_z,TT_z} (\vc{r})
=
\nonumber\\
&&
{\cal A} \Bigl\{
\bigl[
[\chi^{(1)}_\frac{1}{2} \chi^{(2)}_\frac{1}{2} ]_s
\phi_{nlm}(\vc{r}) 
\bigr]_{J=1 , M_z}
[\eta^{(1)}_I \eta^{(2)}_{I'} ]_{T,T_z}
\Bigr\}
,
\end{eqnarray}
where ${\cal A}$ is the antisymmetrization operator.
The coordinate $\vc{r} \equiv \vc{r}_1 - \vc{r}_2 $ is the relative coordinate between the two baryons.
For the deuteron, one relative coordinate is needed to express its wave function.
The wave function is given as a superposition of Gaussian base functions
\begin{eqnarray}
\phi_{nlm}(\vc{r})
&=&
N_{nl } r^l \, e^{-(r/r_n)^2}
Y_{lm}({\widehat {\vc r}})  \;  ,
\end{eqnarray}
with $N_{nl} $ the normalization constant.
The Gaussian range parameters are given in a geometric progression
\begin{eqnarray}
r_n
&=&
r_1 a^{n-1} \qquad \enspace
(n=1 - n_{\rm max}) \; .
\end{eqnarray}

The hamiltonian of the deuteron is given by
\begin{eqnarray}
H
&=&
\sum_{a=1}^2 T_a
+\Delta M
+ V_{NN}
+V_{YN}
\nonumber\\
&&
+\sum_{a=1}^2 T^{(|\Delta S|=1)}_{a}
+{\cal H}_{P\hspace{-.5em}/\, }^{(|\Delta S|=1)}
,
\label{eq:hamil7}
\end{eqnarray}
with the kinetic energy $T$, the nuclear force $V_{NN}$, the hyperon-nucleon potential $V_{YN}$, the strangeness violating weak one-body transition $T^{(|\Delta S|=1)}_{ a}$, and the P-odd meson exchange strangeness violating two-body potential ${\cal H}_{P\hspace{-.5em}/\,}^{(|\Delta S|=1)}$.
The effect of the shift of the baryon mass is incorporated via the mass shift $\Delta M = m_B - m_N$, where $B = N , \Lambda$, or $\Sigma$, depending on the channel chosen with baryons $NB$.
The baryon masses we use are $m_N =939$ MeV, $m_\Lambda =1115$ MeV, and $m_\Sigma =1193$ MeV \cite{pdg}.

\subsection{$NN$ and $YN$ interactions}

In our calculation, we adopt the Argonne $v18$ potential \cite{av18} as the $NN$ interaction.
Our calculation using the Gaussian Expansion Method yields the deuteron binding energy 2.224536 MeV, which reproduces the original result of Ref. \cite{av18} by 5 digits.

For the $YN$ potential of the even parity states, we use the interaction of Ref. \cite{hiyamaYN2}.
This potential was determined so as to simulate the scattering phase shifts given by the $YN$ interaction NSC97f \cite{NSC97f}, and reproduces the binding energies of $^4_\Lambda$H and $^4_\Lambda$He as 2.33 MeV and 2.28 MeV, respectively.
It also yields the $^3_\Lambda$H binding energy $B=0.19$ MeV, which is consistent with the observed data $B=0.13 \pm 0.05$ MeV.
For the odd parity state, we set the $YN$ interaction as vanishing.

\subsection{Quark level weak effective hamiltonian\label{sec:quarklevelweak}}

After the integration of the $W$ boson, the $|\Delta S| =1$ effective hamiltonian of SM is given by
\begin{eqnarray}
&&{\cal H}_{eff} (\mu = m_W)
\nonumber\\
&=&
\frac{G_F}{\sqrt{2}}
\Biggl\{
C_1 (\mu = m_W) [ V_{us}^* V_{ud} Q_1^u + V_{cs}^* V_{cd} Q_1^c ]
\nonumber\\
&& \hspace{3em}
+C_2 (\mu = m_W) [ V_{us}^* V_{ud} Q_2^u + V_{cs}^* V_{cd} Q_2^c ]
\nonumber\\
&& \hspace{4em}
- V_{ts}^* V_{td} \sum_{i=3}^6 C_i (\mu = m_W) Q_i
\Biggr\}
+{\rm h.c.}
,
\label{eq:effhamimw}
\end{eqnarray}
with $V_{qq'}$ the CKM matrix elements and $G_F = 1.16637 \times 10^{-5} {\rm GeV}^{-2}$ the Fermi constant \cite{pdg}.
Here the $\Delta S =-1$ four-quark operators $Q_i$ ($i=1 \sim 6$) are given in the following basis \cite{buras}
\begin{eqnarray}
Q_1^q &=&
\bar s_\alpha \gamma^\mu (1-\gamma_5) q_\beta \cdot \bar q_\beta \gamma_\mu (1-\gamma_5) d_\alpha
,
\label{eq:q1}
\\
Q_2^q &=&
\bar s_\alpha \gamma^\mu (1-\gamma_5) q_\alpha \cdot \bar q_\beta \gamma_\mu (1-\gamma_5) d_\beta
,
\label{eq:q2}
\\
Q_3 &=&
\bar s_\alpha \gamma^\mu (1-\gamma_5) d_\alpha \cdot \sum_q^{N_f} \bar q_\beta \gamma_\mu (1-\gamma_5) q_\beta
,
\label{eq:q3}
\\
Q_4 &=&
\bar s_\alpha \gamma^\mu (1-\gamma_5) d_\beta \cdot \sum_q^{N_f} \bar q_\beta \gamma_\mu (1-\gamma_5) q_\alpha
,
\label{eq:q4}
\\
Q_5 &=&
\bar s_\alpha \gamma^\mu (1-\gamma_5) d_\alpha \cdot \sum_q^{N_f} \bar q_\beta \gamma_\mu (1+\gamma_5) q_\beta
,
\label{eq:q5}
\\
Q_6 &=&
\bar s_\alpha \gamma^\mu (1-\gamma_5) d_\beta \cdot \sum_q^{N_f} \bar q_\beta \gamma_\mu (1+\gamma_5) q_\alpha
,
\label{eq:q6}
\end{eqnarray}
where $\alpha$ and $\beta$ are the color indices of the quarks.

Here it is convenient to Fierz transform the operators $Q_5$ and $Q_6$:
\begin{eqnarray}
Q_5
&=&
\frac{2}{3} \sum_{q=u,d,s} \bar s (1+\gamma_5) q \, \bar q (1-\gamma_5) d 
\nonumber\\
&&
+ 4 \sum_{q=u,d,s} \sum_a \bar s (1+\gamma_5) t_a q \, \bar q (1-\gamma_5) t_a d 
,
\label{eq:q5fierz}
\\
Q_6
&=&
2 \sum_{q=u,d,s} \bar s (1+\gamma_5) q \, \bar q (1-\gamma_5) d 
,
\label{eq:q6fierz}
\end{eqnarray}
where $t_a$ is the generator of the color $SU(3)_c$ group.
The operators $Q_5$ and $Q_6$ are often called QCD penguin operators.

The Wilson coefficients are evolved down to the hadronic scale according to the renormalization group equation at the next-to-next leading logarithmic order \cite{buras,yamanakasmedm}.
Near the hadronic scale $\mu =$1 GeV, the effective hamiltonian can be expressed as
\begin{equation}
{\cal H}_{eff} (\mu)
=
\frac{G_F}{\sqrt{2}} V_{us}^* V_{ud}
\sum_{i=1}^6
[ z_i (\mu) + \tau y_i (\mu)] Q_i (\mu)
+ {\rm h.c.}
,
\label{eq:effhamibelowmc}
\end{equation}
where $\tau \equiv - \frac{V_{ts}^* V_{td}}{V_{us}^* V_{ud}}$.
After renormalization, the Wilson coefficients $y_i$ and $z_i$ ($i=1 \sim 6$) at the hadronic scale $\mu = 1$ GeV are given by
\begin{equation}
{\bf z} (\mu = 1 \, {\rm GeV})
=
\left(
\begin{array}{c}
-0.107 \cr
1.02 \cr
1.76 \times 10^{-5} \cr
-1.39 \times 10^{-2}  \cr
6.37 \times 10^{-3} \cr
-3.45 \times 10^{-3} \cr
\end{array}
\right)
,
\end{equation}
and 
\begin{equation}
{\bf y} (\mu = 1 \, {\rm GeV})
=
\left(
\begin{array}{c}
0 \cr
0 \cr
1.48 \times 10^{-2} \cr
-4.81 \times 10^{-2} \cr
3.22 \times 10^{-3} \cr
-5.69 \times 10^{-2} \cr
\end{array}
\right)
.
\label{eq:zy}
\end{equation}

\subsection{Hyperon-nucleon transition}

The weak one-body hyperon-nucleon transition relevant in this work is given by
\begin{eqnarray}
T^{(|\Delta S|=1)}
&=&
-a_{p \Sigma^+} [ p^\dagger \Sigma^+ ]
-a_{n \Sigma^0} [ n^\dagger \Sigma^0 ]
-a_{p \Lambda} [ n^\dagger \Lambda ]
\nonumber\\
&&
+({\rm h.c.})
.
\label{eq:hyperon-nucleon}
\end{eqnarray}
Here the hyperon and nucleon creation/annihilation operators $\bigl[ \,N^\dagger Y \, \bigr]$ ($Y=\Lambda , \Sigma^0 , \Sigma^+$; $N=p,n$) were explicitly written.
The hyperon-nucleon transition contributing to the nuclear EDM receives the dominant contribution from the $|\Delta S|=1$ four-quark operators $Q_1$ [Eq. (\ref{eq:q1})] and $Q_2$ [Eq. (\ref{eq:q2})], with the strangeness violation by the CKM matrix elements $V_{us} V_{ud}^*$.
The calculation of the coupling constants $a_{p \Sigma^+}$, $a_{n \Sigma^0}$, and $a_{p \Lambda}$ requires the evaluations of the renormalization group evolution of the Wilson coefficients of the $|\Delta S|=1$ four-quark operators and the $|\Delta S|=1$ baryon matrix elements.
We use the result of the renormalization group at the next-to-leading logarithmic approximation to obtain the Wilson coefficients \cite{buras,yamanakasmedm}.
For the baryon matrix elements, we use the result of the calculation in the nonrelativistic quark model \cite{hiyamahyperon-nucleon}.
The detailed expressions of the couplings are given in Appendix \ref{sec:YNtransition}.

\subsection{$|\Delta S | = 1$ P-odd interbaryon force\label{sec:pvbb}}

We now construct the $|\Delta S | = 1$ P-odd interbaryon force by assuming the one-meson (pion, kaon, and eta meson) exchange. 
It is given by combining the $|\Delta S|=1$ P-odd meson-baryon interaction with the standard meson-baryon interaction of the chiral $SU(3)$ lagrangian.
The $|\Delta S|=1$ P-odd meson-baryon interaction is calculated with the use of the vacuum saturation approximation \cite{4-quark1,4-quark2,4-quark3,4-quark4,4-quark5}
\begin{eqnarray}
\langle  B'' \phi | \bar d \gamma_5 d\, \bar s d | B \rangle
&\approx&
\langle \phi | \bar d \gamma_5 d | 0 \rangle \langle B'' | \bar s d | B \rangle
,
\label{eq:vacuumsaturation}
\end{eqnarray}
where the matrix element of the baryons $B$, $B''$ and that of the meson $\phi$ can be factorized.
The $|\Delta S| =1 $ P-odd meson-baryon interaction then receives contribution only from the penguin operators (\ref{eq:q5}) and (\ref{eq:q6}) \cite{yamanakasmedm}.
The detail of the estimation of the couplings using factorization is given in Appendix \ref{sec:P-oddmB}.

Terms contributing to the deuteron EDM are given by \cite{pvcpvhamiltonian1}
\begin{eqnarray}
{\cal H}_{P\hspace{-.5em}/\,}^{(|\Delta S|=1)}
%%%%%%%%%%%%  Terms with local vertex  %%%%%%%%%%%%%%%
&=&
\frac{1}{2m_N} \Biggl\{\,
-g_{\pi NN} \bar g_{\pi^0 \Lambda n} \Bigl[ \Lambda^\dagger n \Bigr]_2 \vc{ \sigma}_1 \cdot \vc{\nabla} {\cal Y}_\pi (r)
\nonumber\\
&& \hspace{2.5em}
-g_{\pi NN} \bar g_{\pi^0 \Sigma^0 n} \Bigl[ {\Sigma^0}^\dagger n \Bigr]_2 \vc{ \sigma}_1 \cdot \vc{\nabla} {\cal Y}_\pi (r)
\nonumber\\
&& \hspace{2.5em}
-g_{\pi NN} \bar g_{\pi^0 \Sigma^+ p} \Bigl[ {\Sigma^+}^\dagger p \Bigr]_1 \vc{ \sigma}_2 \cdot \vc{\nabla} {\cal Y}_\pi (r)
\nonumber\\
&& \hspace{2.5em}
-g_{\eta NN} \bar g_{\eta \Lambda n} \Bigl[ \Lambda^\dagger n \Bigr]_2 \vc{ \sigma}_1 \cdot \vc{\nabla} {\cal Y}_\eta (r)
\nonumber\\
&& \hspace{2.5em}
-g_{\eta NN} \bar g_{\eta \Sigma^0 n} \Bigl[ {\Sigma^0}^\dagger n \Bigr]_2 \vc{ \sigma}_1 \cdot \vc{\nabla} {\cal Y}_\eta (r)
\nonumber\\
&& \hspace{2.5em}
+g_{\eta NN} \bar g_{\eta \Sigma^+ p} \Bigl[ {\Sigma^+}^\dagger p \Bigr]_1 \vc{ \sigma}_2 \cdot \vc{\nabla} {\cal Y}_\eta (r)
\Biggr\}
\nonumber\\
%%%%%%%%%%%%  Kaon exchange  %%%%%%%%%%%%%%%
&&
-\frac{g_{K N \Lambda} \bar g_{\bar K^0 p p}}{4 \mu_{N \Lambda}}
\Bigl[ \Lambda^\dagger n \Bigr]_2 \vc{ \sigma}_2 \cdot \vc{\nabla} {\cal Y}_K (r)
\nonumber\\
&& 
+\frac{ g_{K N \Sigma} \bar g_{\bar K^0 p p} }{4\sqrt{2} \mu_{N \Sigma}}
\Bigl[ {\Sigma^0}^\dagger n \Bigr]_2 \vc{ \sigma}_2 \cdot \vc{\nabla} {\cal Y}_K (r)
\nonumber\\
&& 
+\frac{ g_{K N \Sigma} \bar g_{\bar K^0 n n} }{4 \mu_{N \Sigma}}
\Bigl[ {\Sigma^+}^\dagger p \Bigr]_1 \vc{ \sigma}_1 \cdot \vc{\nabla} {\cal Y}_K (r)
\nonumber\\
&&
+(1\leftrightarrow 2)
\nonumber\\
&&
+({\rm h.c.})
,
\label{eq:PCVintdeuteron}
\end{eqnarray}
where $\vc{\sigma}_i$ denotes the spin matrix.
The indices 1 and 2 of each operator label the baryon with positive charge and the neutral baryon, respectively.
In this work, we only consider the charge neutral meson exchange interactions, since the charged meson exchange requires the isospin breaking which is a small effect.
The kaon exchange terms have a dependence on the reduced masses $\mu_{N \Lambda} \equiv \frac{m_N m_\Lambda}{m_N + m_\Lambda} $ and $\mu_{N \Sigma} \equiv \frac{m_N m_\Sigma}{m_N + m_\Sigma} $.
The nonlocal terms relevant in the kaon exchange interaction were neglected.
For the derivation of the $|\Delta S|=1$ P-odd interbaryon force in the one-meson exchange model, see Appendix \ref{sec:P-oddBB}.
All terms in the above equation involve the derivative of the Yukawa function $\vc{\nabla} {\cal Y}_X (r)$ which can be rewritten as
\begin{equation}
\vc{\nabla} {\cal Y}_X (r) = -\frac{m_X}{4\pi} \frac{e^{-m_X r }}{r} \left( 1+ \frac{1}{m_X r} \right) \hat{\vc{r}}
\ ,
\end{equation}
where $\hat{\vc{r}}$ is the unit vector pointing from the baryon 1 to the baryon 2.
The P-even meson-baryon couplings are given as $g_{\pi NN}  =  \frac{g_A m_N}{f_\pi}= 12.9$, $g_{\eta NN} = 2.24$ \cite{tiator}, $g_{K\Lambda N}=\frac{m_N+m_\Lambda}{2\sqrt{3} f_\pi } (D+3F) \approx 13.6$ and $g_{K\Sigma N } = \frac{m_N +m_\Sigma}{\sqrt{2} f_\pi } (F-D) \approx -6.0$. 
The last two couplings were derived from the on-shell approximation of the leading terms of the chiral lagrangian \cite{hisano,faessler,fuyuto,rpvlinearprogramming}.

In the context of the deuteron EDM, the $|\Delta S | = 1$ meson-baryon interaction mainly receives contribution from the Penguin process, involving the CKM matrix elements $V_{td} V_{ts}^*$.
It is important to note that the combination of these CKM matrix elements with those of the one-body hyperon-nucleon transition seen previously makes the Jarlskog invariant \cite{jarlskog}, the leading CP violating constant in the SM.
The meson-baryon matrix elements were calculated in the factorization approach, where the vacuum saturation approximation \cite{vacuumsaturation,pvcpvhamiltonian1,factorizationedm1,factorizationedm6,rpvlinearprogramming} was applied to the meson-baryon matrix elements of the $|\Delta S | = 1$ four-quark operator \cite{yamanakasmedm}.
The renormalization group evolution of the Wilson coefficients were calculated in the next-to-leading logarithmic approximation \cite{buras,yamanakasmedm}.
The derivation of the P-odd coupling constants are given in Appendix \ref{sec:P-oddmB}.
We have to note that this step involves the largest theoretical uncertainty of this work.

Here we must note that the $|\Delta S|=1$ P-odd meson-baryon interactions generated by $V_{us} V_{ud}^*$, together with the hyperon-nucleon transition generated by $V_{td} V_{ts}^*$, have smaller effects.
This is because the contribution of $V_{td} V_{ts}^*$ to the hyperon-nucleon transition is subleading, due to the small Wilson coefficients of penguin operators \cite{buras}, and also because the tree level $W$ boson exchange diagram with $V_{us} V_{ud}^*$ does not contribute to the $|\Delta S|=1$ P-odd meson-baryon interaction in the factorization approach.
The contributions of $V_{us} V_{ud}^*$ and $V_{td} V_{ts}^*$ to the effective hadron level interactions are therefore distinct, and no double counting occurs \cite{yamanakasmedm}.

%%%%%%%%%% Definition of EDM %%%%%%%%%%%%%%%

\section{The electric dipole moment\label{sec:edm}}

In the SM, the nuclear EDM has two leading sources.
The first one is the intrinsic nucleon EDM, which has been evaluated in many previous works \cite{stetcu,song,bsaisou,bsaisou2,yamanakanuclearedm}.
The nucleon EDM in the SM is estimated to be around $O(10^{-32})e$ cm for the long distance contribution \cite{seng}.
The deuteron EDM is related to the nucleon EDM by \cite{yamanakanuclearedm,bsaisou2}
\begin{equation}
d_{d}^{\rm (Nedm)}
\approx
0.91 (d_n +d_p)
.
\label{eq:edmintrinsic}
\end{equation}
The coefficient was given by the calculation using the Argonne $v18$ potential \cite{av18}.
This suppression of the nucleon EDM effect is due to the $d$-wave component of the deuteron.
The nucleon EDM contribution to the deuteron EDM is not enhanced since the deuteron is a nonrelativistic system \cite{sandars}.
Recently, additional suppression due to the polarization of the nuclear system by the neutron EDM was also pointed \cite{sinoue}.

The short distance contributions are composed of the single quark EDM, the chromo-EDM \cite{czarnecki}, the Weinberg operator \cite{weinbergop,smweinbergop}, and the loop-less dimension-six operator process \cite{mannel}.
The single quark EDM/chromo-EDM contribution is known to be small, of order $10^{-35}e$ cm.
Moreover, the renormalization group evolution \cite{tensorrenormalization,degrassi,yang} and the nucleon tensor charge \cite{yamanakasde1,pitschmann,etm3,bhattacharya1,bhattacharya2,bhattacharya3,jlqcd4}, the coefficient relating the quark EDM to the nucleon EDM, bring additional suppressions.
The effect of the Weinberg operator on the nucleon EDM is known to be even smaller, of order $10^{-40}e$ cm \cite{smweinbergop,pospelovweinbergop}.
These contributions can therefore be neglected.
The loop-less dimension-six operator effect was estimated to be $O(10^{-31})e$ cm, but this value should be viewed as an upper limit, since the gluon dressing effect may reduce the polarization of the nucleon \cite{yamanakasde1,yamanakasde2}.
For those reasons, we do not consider the effect of the nucleon EDM further.

We now point to the second contribution, the polarization of the nuclear system by the P, CP-odd two-body interaction (P, CP-odd nuclear force).
The effect of nuclear polarization is the main target of our discussion, since the dynamical effect of hypernuclei enters in the evaluation of the deuteron EDM through the $|\Delta S|=1$ weak interaction.
The polarization contribution of the P, CP-odd nuclear force to the deuteron EDM is given by
\begin{eqnarray}
d_{d}^{\rm (pol)} 
&=&
\sum_{i=1}^{2}
\langle \, \tilde d \, |\, e Q_i \, \vc{r}_{iz} \, | \, \tilde d \, \rangle
,
\end{eqnarray}
where $|\, \tilde d\, \rangle$ is the polarized (in the $z$-axis) deuteron wave function, and $e Q_i$ is the charge of the $i$'th particle.
The $z$-component of the coordinate of the constituent nucleon in the nuclear center of mass frame is denoted by $\vc{r}_{iz}$.
This permanent polarization effect is realized through the mixing of opposite parity states.

\begin{figure}[htb]
\begin{center}
\includegraphics[width=6.4cm]{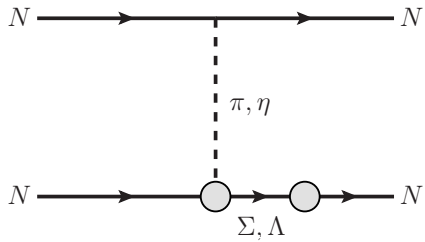}
\caption{\label{fig:CPVNN}
Example of a diagram contributing to the P, CP-odd nuclear force in the SM, evaluated in the previous work \cite{yamanakasmedm}.
The intermediate hyperon state is integrated out.
}
\end{center}
\end{figure}

\begin{figure*}[htb]
\begin{center}
\includegraphics[width=14cm]{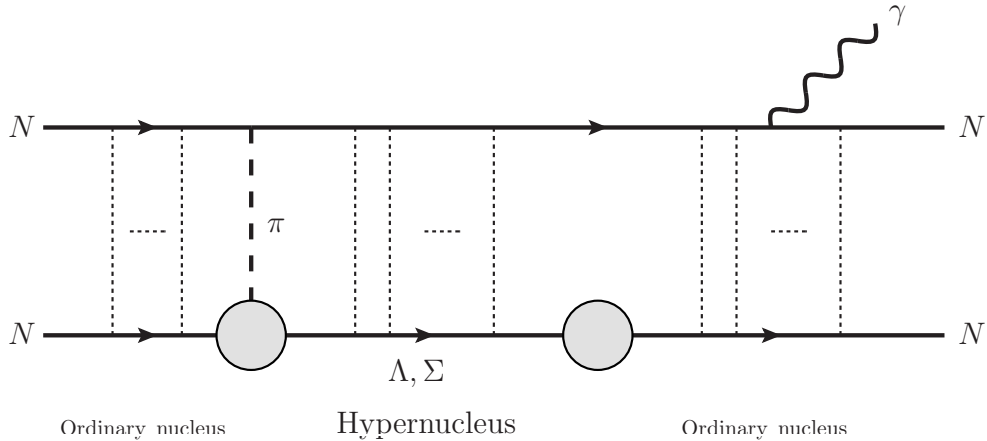}
\caption{\label{fig:deuteron_EDM_cc}
Coupled channel contribution to the nuclear EDM in the SM.
The thick dashed line denotes the P-odd $|\Delta S | = 1$ one-meson exchange interaction, and the thin dashed lines denote the $NN$ and $YN$ interactions.
The photon insertion (wavy line) acts as the EDM operator probing the polarization of the deuteron.
The nucleus-hypernucleus mixing occurs due to the $|\Delta S | = 1$ weak interaction (grey blobs).
}
\end{center}
\end{figure*}

In a previous work, we have calculated the SM contribution to the nuclear EDM by deriving the P, CP-odd nuclear force made of the combination of the $|\Delta S|=1$ interbaryon interaction with one light meson exchange and the one-body hyperon-nucleon transition (see Fig. \ref{fig:CPVNN}) \cite{yamanakasmedm}.
This procedure assumes that the hyperon created by the $|\Delta S|=1$ interbaryon interaction is immediately annihilated by the hyperon-nucleon transition (or vice versa), and neglects the dynamical effect of the interaction between the intermediate hyperon and other nucleons, as well as the polarization of the intermediate hypernuclear state.
As the weak interacting process does not occur often, the hyperon ``survives for a very long off-shell time'' in the nucleus, and this effect  should not be neglected.
This contribution is described by the mixing of an ordinary nucleus with the hypernucleus, with the $NN - \Lambda N - \Sigma N$ coupled channels (see Fig. \ref{fig:deuteron_EDM_cc}).
In this work, we calculate the deuteron EDM in the SM by taking into account the $NN - \Lambda N - \Sigma N$ channel coupling.

%%%%%%%%%%%%%%   Result and analysis   %%%%%%%%%%%%%%%

\section{Result and analysis\label{sec:analysis}}

\begin{figure}[htb]
\begin{center}
\includegraphics[width=8cm]{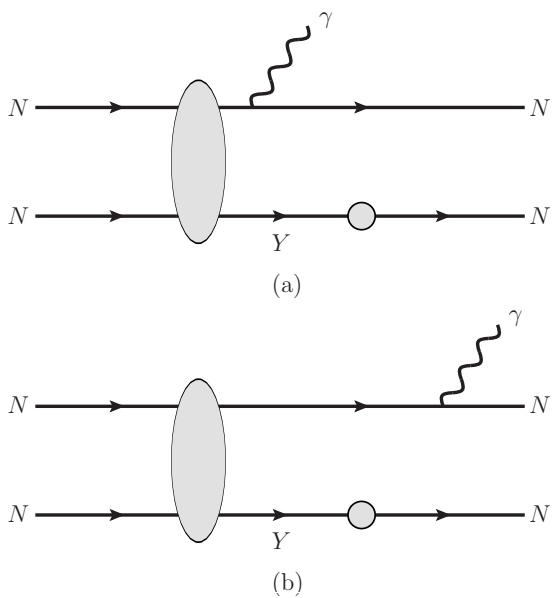}
\caption{\label{fig:YN_polarization}
Schematic diagrams showing the polarizations of the intermediate hypernuclear state (a) and the ordinary nuclear state (b) in the $NN - \Lambda N - \Sigma N$ channel couplings.
}
\end{center}
\end{figure}

From our calculation, the resulting EDM of the deuteron is
\begin{equation}
d_d = 2.8 \times 10^{-31} e \, {\rm cm}
.
\label{eq:treeleveledm}
\end{equation}
The detailed breakdown is shown in Table \ref{table:result}.
We see that the pion exchange contribution is dominant.
By comparing the result of this work (fifth column of Table \ref{table:result}) with that of the previous work (second column of Table \ref{table:result}), a deviation less than 10\% can be observed.
This shift is mainly due to the additional effect from the polarization of the intermediate hypernuclear states [see Fig. \ref{fig:YN_polarization} (a)].

We also see that for the pion exchange, the contribution from the free $YN$ intermediate state without taking into account the $YN$ polarization (third column of Table \ref{table:result}) agrees with the result of the previous work \cite{yamanakasmedm} within 2\%.
This means that this nonrelativistic coupled channel analysis agrees with the P, CP-odd nuclear force with integrated out hyperons, used in the previous work \cite{yamanakasmedm}.
As this approximation is using the on-shell Dirac equation of the nucleon to integrate out the intermediate hyperon, this slight discrepancy can be thought as the relativistic effect of the hyperon-nucleon transition.
For the pion exchange, the relativistic effect is therefore small.

Another important feature is that the dynamical effect from the hyperon-nucleon potential is subleading, with the effect of $\Lambda N $ interactions slightly larger than that of $\Sigma N$ interactions (see Table \ref{table:result}).
This means that, at least for the deuteron, the hyperon and the nucleon of the $S =-1$ intermediate state can be considered as free.

The contribution from eta and kaon exchange weak interactions are subdominant in the deuteron EDM, less than 10\% of the total deuteron EDM.
Although being subleading, the analysis of heavier meson exchange brings us important knowledge on the relativistic effect.
By comparing with the previous work, we see here deviations by about 20\% for the eta exchange effect, and about 10\% for the kaon one.
In particular, we observe that the contribution from the free $YN$ intermediate state without taking into account the $YN$ polarization (third column of Table \ref{table:result}) is significantly shifted from that of the previous work (second column of Table \ref{table:result}).
We interpret this difference as the relativistic effect, which should be more important for the eta and kaon exchange processes than that with pions.

Here we should explain in detail the relativistic effect relevant in the hyperon-nucleon transition.
In the previous work \cite{yamanakasmedm}, the intermediate state with a hyperon has been integrated out by using the on-shell Dirac equation.
In this work, however, we take into account the intermediate states with hyperons, and to do so we used the nonrelativistic Schr\"{o}dinger equation with coupled channels.
We must then use the nonrelativistic hyperon-nucleon transition matrix elements, so our current calculation does not include the nonrelativistic effect.
The difference between the current results and the previous one is therefore a probe of nonrelativistic effect of the hyperon-nucleon transition.
The result is consistent with this interpretation, since the current result and that of the previous work have larger discrepancy when the mass of the exchanged meson increases (the increase of the meson mass means that the momentum transfer becomes larger for observables which are generated by this meson exchange process).

For the kaon exchange contribution, we additionally have theoretical uncertainty due to the neglect of nonlocal interactions (see Appendix \ref{sec:P-oddBB}).
This uncertainty is at the level of $O( |m_Y -m_N| /  m_N )$.
The kaon exchange effect is less deviated than that of the eta meson exchange in comparing the hyperon integration (result of previous work) and this nonrelativistic coupled channel analysis.
This difference is because the kaon is lighter than the eta meson so that the relativistic effect is less important.
The shift due to the relativistic effect as well as the uncertainty due to the omission of nonlocal interactions are difficult systematics to control in the framework used in this work.
The systematics for the kaon and eta exchange processes is fortunately comparable to the error bar of the relativistic effect of pion exchange contribution.
We can thus conclude that the relativistic correction is of order of few percents.
We can also expect that the shift for the $^3$He or $^3$H EDMs is of the same order, since the kinetic energy of the nucleon inside those systems are comparable to that for the deuteron and the relativistic corrections are consequently of the same order.

We must also note that there is an uncertainty in the realistic nuclear potential at the short distance.
The results of the calculations of the EDMs of the deuteron, $^3$He and $^3$H within several realistic nuclear forces are in good agreement for the pion exchange P, CP-odd $NN$ interaction.
However, the contributions from the P, CP-odd $NN$ interaction with the exchange of heavier mesons show discrepancies depending on the realistic nuclear force chosen \cite{liu,song}.
We therefore have additional theoretical uncertainty for the deuteron EDM generated by the kaon and eta meson exchanges.

The most important modification due to the consideration of the coupled $NN - \Lambda N - \Sigma N$ channels is therefore the polarization effect of the intermediate hypernuclear state [see Fig. \ref{fig:YN_polarization} (a)].
We have been possible to determine a so far unknown nuclear level source of systematics, and confirm that it does not change the order of magnitude of the deuteron EDM.
All those examinations and statements have to be confirmed in future works for $^3$He and $^3$H, which should present similar physics.

What about $p$-shell nuclei and beyond?
For those systems, we must note that the Pauli exclusion principle is important.
On the other hand, the EDM of $^{13}$C was studied in a recent work from the point-of-view of the nuclear structure, and it was found that the change of the nuclear structure significantly suppresses the EDM \cite{13cedm}.
For heavier nuclei with the relevant Pauli exclusion effect, their corresponding hypernuclei have significantly different structures \cite{motoba1,motoba2,bando,yamadahyp,spectroscopyhypernuclei,tanida,kohri,isaka,hagino,hiyamasu(3),schulze}, so that their EDM may be suppressed.
The EDMs of $^6$Li and $^9$Be in the SM may be a good target to unveil the modification due to Pauli exclusion effect.

\begin{table*}[hbt]
\caption{
The breakdown of the deuteron EDM in the SM with $NN - \Lambda N - \Sigma N$ coupled channels.
}
\vskip 0.2cm
\begin{tabular}{ccccc}
\hline
\hline
Weak $|\Delta S|=1$ interaction, \ \ & Integrated out hyperons\footnote{We have corrected the previous calculation by changing the sign of the isovector eta exchange P, CP-odd nuclear coupling $\bar G_\eta^{(1')}$ and that of the isovector kaon exchange P, CP-odd nuclear coupling $\bar G_K^{(1)}$, which were wrong in Ref. \cite{yamanakasmedm}.
The value adopted for the CP-even pion-nucleon coupling is $g_{\pi NN}=12.9$.} & Free $YN$, & Free $YN$, & Interacting $YN$, \\
exchanged meson & (previous work \cite{yamanakasmedm}) & no $YN$ polarization & with $YN$ polarization & with $YN$ polarization \\
\hline
\vspace {-3 mm} \\
$\Lambda N -NN , \pi $ & $-9.85 \times 10^{-32} e$ cm & $-9.70 \times 10^{-32} e$ cm  & $-8.54 \times 10^{-32} e$ cm & $-8.25 \times 10^{-32} e$ cm \\
$\Sigma N -NN , \pi $ & $3.37 \times 10^{-31} e$ cm & $3.43 \times 10^{-31} e$ cm  & $3.50 \times 10^{-31} e$ cm & $3.50 \times 10^{-31} e$ cm \\
\hline
\vspace {-3 mm} \\
$\Lambda N -NN , \eta $ & $2.67 \times 10^{-33} e$ cm & $3.23 \times 10^{-33} e$ cm  & $2.72 \times 10^{-33} e$ cm & $2.53 \times 10^{-33} e$ cm \\
$\Sigma N  -NN , \eta $ & $3.05 \times 10^{-33} e$ cm & $3.73 \times 10^{-33} e$ cm  & $4.95 \times 10^{-33} e$ cm & $4.89 \times 10^{-33} e$ cm \\
\hline
\vspace {-3 mm} \\
$\Lambda N  -NN, K^0 $ & $4.40 \times 10^{-32} e$ cm & $4.76 \times 10^{-32} e$ cm  & $4.02 \times 10^{-32} e$ cm & $3.76 \times 10^{-32} e$ cm \\
$\Sigma N  -NN, K^0 $ & $-2.48 \times 10^{-32} e$ cm & $-2.64 \times 10^{-32} e$ cm  & $-3.34 \times 10^{-32} e$ cm & $-3.31 \times 10^{-32} e$ cm \\
\hline
\vspace {-3 mm} \\
Total & $2.63 \times 10^{-31} e$ cm  & $2.75 \times 10^{-31} e$ cm  & $2.79 \times 10^{-31} e$ cm & $2.79 \times 10^{-31} e$ cm \\
\hline
\end{tabular}
\label{table:result}
\end{table*}

Let us now estimate the theoretical uncertainty of this calculation.
The largest systematics is due to the $|\Delta S| =1$ P-odd interbaryon potential.
In our work, we have modeled the $NN \to YN$ interaction by the one meson exchange effect, with the $|\Delta S|=1$ interactions calculated using the vacuum saturation approximation.
The glunoic correction to the CP-odd meson-baryon interaction generated by a CP-odd four-quark interaction is typically $O(100\%)$ in the large $N_c$ analysis \cite{thooft,manohar,samart}.
In the case of $|\Delta S|=1$ processes, however, we have a strange quark which forbids the exchange of quarks between mesons and baryons, reducing higher order gluonic corrections.
Moreover, the $|\Delta S|=1$ four-quark interaction relevant in our discussion does not involve up-quarks, so that the possibility to draw gluon exchange diagrams with nucleons are additionally suppressed.
We therefore estimate the theoretical uncertainty due to the factorization to be $2/N_c \sim 67 \%$.

An additional important source of theoretical uncertainty is due to higher order terms of $SU(3)$ chiral perturbation theory, which bring $O(m_K^2/\Lambda^2) \sim O(m_\eta^2/\Lambda^2) \sim 25\%$ ($\Lambda \sim 1$ GeV) to the $|\Delta S| =1$ interbaryon force.
We also consider the QCD level uncertainty, originating from the renormalization group evolution, quark masses, etc.
As all QCD level quantities have been renormalized in the next-to-leading logarithmic order \cite{buras}, it is adequate to estimate the error bar by $\alpha_s^2 (\mu = 1\, {\rm GeV}) \sim 0.2$.
We also include the uncertainty due to the intrinsic nucleon EDM [see Eq. (\ref{eq:edmintrinsic})].
Here we consider that the nucleon EDM has null central value, and adopt the uncertainty $\delta d_d^{\rm (Nedm)} \sim 1 \times 10^{-31}e$ cm, suggested by previous studies \cite{mannel,seng}.
To sum up, the total error bar is estimated as
\begin{eqnarray}
\delta d_d 
&=& 
\sqrt{
d_d^2 \bigl[
\bigl( 2/N_c \bigr)^2
+ ( m_K^2/\Lambda^2 )^2
+ ( \alpha_s^2 )^2
\bigr]
+ \delta d_d^{\rm (Nedm)2}
}
\nonumber\\
&\approx &
0.8 d_d
.
\label{eq:totaledmerror}
\end{eqnarray}
This gives the final result
\begin{equation}
d_d = 
(2.8 \pm 2.3) \times 10^{-31} e \, {\rm cm}
.
\end{equation}

There are currently no accurate data to determine the $|\Delta S|=1$ P-odd meson-baryon vertex.
However, we have to note that the error bar due to the hadron physics is independent of the nuclear level calculation of this work. 
The theoretical uncertainty due to the $|\Delta S| =1$ P-odd meson-baryon matrix elements is due to the hadron physics, but the deviation of the EDM of the deuteron due to the consideration of $NN-YN$ coupled channels is a nuclear level phenomenon.
Those two effects are well separated in energy scale, so the enhancement of 10\% of the deuteron EDM is not affected by the theoretical uncertainty of the $|\Delta S| =1$ P-odd meson-baryon interaction, which is just a constant in our work. 
This determination of the nuclear level systematic error is important in the analysis of the deuteron EDM.

As ways to improve the $YN-NN$ coupling potential, we have the phenomenological derivation of the low energy constants of the nonleptonic weak decay of hypernuclei \cite{inoue1,inoue2,sasaki1,sasaki2,perez1,perez2,perez3}.
The determination of the low energy constants ($|\Delta S|=1$ meson-baryon couplings) for $|\Delta S| =1$ four-quark operators in the effective field theory may improve the $|\Delta S| =1$ P-odd interbaryon potential.
The ultimate theoretical approach to determine them is the lattice QCD simulation.
There it is now possible to extract the nuclear potentials from first principle \cite{hal1,hal2}.
If this technique is applicable to the $|\Delta S| =1$ P-odd interbaryon interaction, significant progress is expected in the evaluation of the nuclear EDM in the SM.

%%%%%%%%%%%%%%   Summary   %%%%%%%%%%%%%%%

\section{Summary\label{sec:summary}}

In this paper we have calculated the EDM of the deuteron in the SM by considering the $NN - \Lambda N - \Sigma N$ channel coupling with the Gaussian Expansion Method.
The $|\Delta S| =1$ P-odd meson-baryon interaction has been calculated in the factorization approach.
We have found that the polarization effect in $YN$ channels modifies by less than 10\% the total deuteron EDM.
The modification due to the $YN$ interaction is found to be subleading.
The deuteron EDM in SM is below the experimental sensitivity of the planned experiment using storage rings ($\sim 10^{-29} e$ cm).
This analysis was important in the determination of the nuclear level systematics of the SM contribution to the deuteron EDM.

We expect that similar analyses can be performed for other light nuclei, such as the $^3$He, $^3$H, $^6$Li, $^9$Be, etc.
For heavier nuclei, the Pauli exclusion principle becomes relevant, so that the modification due to the transition of a nucleon to a hyperon could potentially be important.
The study of the SM contribution to the nuclear EDM therefore becomes strongly dependent on the physics of the structure of hypernuclei.

\begin{acknowledgments}
The author thanks Emiko Hiyama, Hajime Togashi, and Bira van Kolck for useful discussions and comments.
This work is supported by the RIKEN iTHES Project.
\end{acknowledgments}

\appendix

\section{Hyperon-nucleon transition\label{sec:YNtransition}}

Here we give the weak coupling constants of hyperon-nucleon transition.
\begin{eqnarray}
a_{p \Sigma^+} 
&=&
|V_{us} V_{ud}| \frac{G_F}{\sqrt{2}} 
\Biggl\{
(z_1 -z_2)
\langle p | Q_2^{NR} | \Sigma^+ \rangle
\nonumber\\
&&-
\Biggl[ \frac{2}{3} y_5 +2 y_6 \Biggr]
\langle p | \bar d s | \Sigma^+ \rangle
\langle 0 | \bar dd + \bar ss | 0 \rangle
\Biggr\}
,
\end{eqnarray}
\begin{eqnarray}
a_{n \Sigma^0} 
&=&
|V_{us} V_{ud}| \frac{G_F}{\sqrt{2}} 
\Biggl\{
(z_1 -z_2)
\langle n | Q_2^{NR} | \Sigma^0 \rangle
\nonumber\\
&&-
\Biggl[ \frac{2}{3} y_5 +2 y_6 \Biggr]
\langle n | \bar d s | \Sigma^0 \rangle
\langle 0 | \bar dd + \bar ss | 0 \rangle
\Biggl\}
,
\end{eqnarray}
\begin{eqnarray}
a_{n \Lambda} 
&=&
|V_{us} V_{ud}| \frac{G_F}{\sqrt{2}} 
\Biggl\{
(z_1 -z_2)
\langle n | Q_2^{NR} | \Lambda \rangle
\nonumber\\
&&-
\Biggl[ \frac{2}{3} y_5 +2 y_6 \Biggr]
\langle n | \bar d s | \Lambda \rangle
\langle 0 | \bar dd + \bar ss | 0 \rangle
\Biggr\}
.
\end{eqnarray}
Note that the phase was fixed so as to obtain real couplings.
The operator $Q_2^{NR}$ is the nonrelativistic reduction of the four-quark operator $Q_2 \equiv \bar s \gamma^\mu (1-\gamma_5) q \cdot \bar q \gamma_\mu (1-\gamma_5) d$.
The hyperon-nucleon transition matrix elements are given by \cite{hiyamahyperon-nucleon}
\begin{eqnarray}
\langle p | Q_2^{NR} | \Sigma^+ \rangle
&=&
2.76 \times 10^{-2} {\rm GeV}^3
,
\\
\langle n | Q_2^{NR} | \Sigma^0 \rangle
&=&
-1.95 \times 10^{-2} {\rm GeV}^3
,
\\
\langle n | Q_2^{NR} | \Lambda \rangle
&=&
-9.65 \times 10^{-3} {\rm GeV}^3
.
\end{eqnarray}
For the baryon scalar density matrices, we have adopted
\begin{eqnarray}
\langle n | \bar d s | \Lambda \rangle
&\approx&
\sqrt{\frac{3}{2}} \cdot \frac{m_N -m_\Lambda}{m_s}
\approx
-1.80
,
\label{eq:sdlamneu}
\\
\langle n | \bar d s | \Sigma^0 \rangle
&\approx&
\frac{ m_\Sigma - m_N}{\sqrt{2} m_s}
\approx
\ \ 1.50
,
\label{eq:sdsigneu}
\\
\langle p | \bar d s | \Sigma^+ \rangle
&\approx&
\frac{m_N - m_\Sigma}{m_s}
\approx
-2.12
\label{eq:sdsigpro}
,
\end{eqnarray}
where the renormalization scale is 1 GeV.
The strange quark mass $m_s = 120$ MeV was obtained through the two-loop level renormalization group evolution.
The chiral condensates renormalized at the same scale are
\begin{equation}
\langle 0 | \bar dd  | 0 \rangle
\approx
\langle 0 | \bar ss | 0 \rangle
\approx
\langle 0 | \bar qq  | 0 \rangle
\approx
-(265\, {\rm MeV})^3
.
\label{eq:chiralcondensate}
\end{equation}
The Wilson coefficients $z_1$ and $z_2$, $y_5$, and $y_6$ are renormalized at $\mu = 1$ GeV.
After the renormalization group evolution in the next-to-leading logarithmic approximation \cite{buras,yamanakasmedm}, we obtain
\begin{eqnarray}
z_1
&=&
-0.0457
,
\\
z_2
&=&
1.23
,
\\
y_5
&=&
9.65 \times 10^{-3}
,
\\
y_6
&=&
-8.40 \times 10^{-2}
.
\end{eqnarray}

\section{$|\Delta S|=1$ P-odd baryon-baryon interaction\label{sec:P-oddBB}}

\begin{figure}[htb]
\begin{center}
\includegraphics[width=6.4cm]{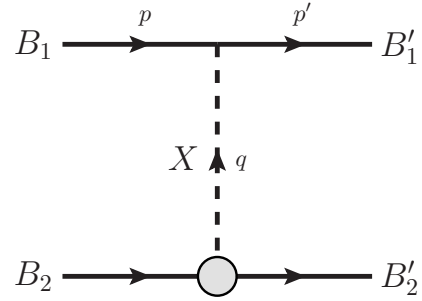}
\caption{\label{fig:CPVFT}
Diagrammatic representation of the one-meson exchange P-odd interbaryon interaction.
The grey blob denotes the P-odd meson-nucleon interaction.
}
\end{center}
\end{figure}

The one-(pseudoscalar)meson exchange P-odd interbaryon interaction is made by combining the standard P-even meson-baryon interaction
\begin{equation}
{\cal L}_{P} = 
g m \bar B'_1 i \gamma_5 B_1 , 
\end{equation}
and the P-odd meson-baryon interaction.
\begin{equation}
{\cal L}_{P\hspace{-0.5em}/} = 
g' m \bar B'_2 B_2.
\end{equation}
The one-meson exchange process is depicted in Fig. \ref{fig:CPVFT}.

The effective lagrangian of the one-meson exchange P-odd baryon-baryon interaction can be written as
\begin{equation}
{\cal L}_{BB} =
\frac{-gg'}{q^2 - m_X^2} \bar B'_1 i \gamma_5 B_1 \cdot \bar B'_2 B_2
,
\label{eq:LBB}
\end{equation}
where $m_X$ is mass of the meson $X$.
The corresponding hamiltonian in the nonrelativistic limit is
\begin{eqnarray}
{\cal H}_{BB} 
&=&
\frac{gg'}{q^2 - m_X^2} \bar B'_1 i \gamma_5 B_1 \cdot \bar B'_2 B_2
\nonumber\\
&\approx &
\frac{1}{2 m_B} \cdot \frac{i gg'}{|\vc{q}|^2 + m_X^2} \chi_{B'_1}^\dagger 
\vc{\sigma} \cdot \Biggl[ 
\frac{\vc{p'}}{2m_{B_1'}}
-\frac{\vc{p}}{2m_{B_1}}
\Biggr]
\chi_{B_1}
\nonumber\\
&& \hspace{12em}
\cdot \chi_{B'_2}^\dagger \chi_{B_2}
,
\end{eqnarray}
where $\vc{p}$ and $\vc{p}'$ are the momenta carried by the baryon $B_1$ and $B_1'$, respectively (see Fig. \ref{fig:CPVFT} for the definition).
The nonrelativistic spinor of the baryon $B$ is denoted by $\chi_B$.

The dependence of the above equation on the momenta carried by $B_1$ and $B_1'$ will in the generic case bring a nonlocal term:
\begin{eqnarray}
\frac{\vc{p'}}{2m_{B_1'}}
-\frac{\vc{p}}{2m_{B_1}}
&=&
\frac{1}{4 \mu_1} (\vc{p}' -\vc{p})
\nonumber\\
&&
+\frac{1}{4} \frac{m_{B_1} -m_{B_1'} }{m_{B_1} m_{B_1'}} (\vc{p}' +\vc{p})
,
\end{eqnarray}
where $\mu_1 \equiv \frac{m_{B_1'} +m_{B_1}}{m_{B_1'} m_{B_1}}$ is the reduced mass.
This second term will generate a nonlocal interaction after Fourier transform.
This term is however suppressed by the mass difference $m_{B_1} -m_{B_1'}$, so we neglect it in this work, and we keep only the term depending on the momentum exchange.

The dependence of the hamiltonian on the exchanged momentum $\vc{q} \equiv \vc{p}' -\vc{p}$ can be Fourier transformed as follows:
\begin{eqnarray}
\int \frac{d^3 q}{(2\pi )^3}
\frac{\vc{q} e^{+i \vc{q} \cdot \vc{r} }}{|\vc{q}|^2 +m_X^2}
&=&
-i \vc{\nabla} 
\int \frac{d^3 q}{(2\pi )^3}
\frac{ e^{+i \vc{q} \cdot \vc{r} }}{|\vc{q}|^2 +m_X^2}
\nonumber\\
&=&
\frac{-i \vc{\nabla} }{4\pi} \frac{e^{-m_X |\vc{r}|}}{|\vc{r}|}
,
\end{eqnarray}
where $\vc{r} \equiv \vc{r}_1 - \vc{r}_2$ with $\vc{r}_1$ and $\vc{r}_2$ the coordinates of the baryons $B_1$ (or $B_1'$) and $B_2$ (or $B_2'$), respectively, and $\vc{\nabla}$ is the derivative with respect to $\vc{r}$.

The hamiltonian of the one-meson exchange P-odd interbaryon interaction expressed in the coordinate space is then
\begin{equation}
{\cal H}_{BB}
\approx
\frac{ gg'}{4 \mu_1} 
(\vc{\sigma}_1 \cdot \vc{\nabla}) \frac{e^{-m_X |\vc{r}|}}{4\pi |\vc{r}|}
,
\end{equation}
where $\vc{\sigma}_1$ is the spin matrix acting on the baryon $B_1$.

\section{Direct contribution to the $|\Delta S|=1$ P-odd meson-baryon interaction\label{sec:P-oddmB}}

The $|\Delta S| =1$ P-odd meson-baryon couplings used in this work are obtained by using the vacuum saturation approximation \cite{vacuumsaturation} to the meson-baryon matrix elements with $|\Delta S| =1$ four-quark operators.
They are given by
\begin{eqnarray}
\bar g_{\pi^0 \Lambda n}
&\approx&
-G_y 
\langle \pi^0 | \bar d \gamma_5 d | 0 \rangle \langle \Lambda | \bar s d | n \rangle
,
\\
\bar g_{\pi^0 \Sigma^0 n}
&\approx&
-G_y 
\langle \pi^0 | \bar d \gamma_5 d | 0 \rangle \langle \Sigma^0 | \bar s d | n \rangle
,
\\
\bar g_{\pi^0 \Sigma^+ p}
&\approx&
-G_y 
\langle \pi^0  | \bar d \gamma_5 d | 0 \rangle \langle \Sigma^+ |  \bar s d | p \rangle
,
\\
\bar g_{\eta \Lambda n}
&\approx&
G_y 
\langle \eta | \bar s \gamma_5 s - \bar d \gamma_5 d | 0 \rangle \langle \Lambda | \bar s d | n \rangle
,
\\
\bar g_{\eta \Sigma^0 n} 
&\approx&
G_y 
\langle \eta | \bar s \gamma_5 s - \bar d \gamma_5 d | 0 \rangle \langle \Sigma^0 | \bar s d | n \rangle
,
\\
\bar g_{\eta \Sigma^+ p}
&\approx&
G_y 
\langle \eta  | \bar s \gamma_5 s - \bar d \gamma_5 d | 0 \rangle \langle \Sigma^+ |  \bar s d | p \rangle
,
\\
\bar g_{\bar K^0 p p} 
&\approx&
G_y 
\langle \bar K^0 | \bar s \gamma_5 d | 0 \rangle \langle p |  \bar d d |p \rangle
,
%\\
\end{eqnarray}
\begin{eqnarray}
\bar g_{\bar K^0 n n} 
&\approx&
G_y 
\langle \bar K^0  | \bar s \gamma_5 d | 0 \rangle \langle n |  \bar d d | n \rangle
,
\end{eqnarray}
where $G_y \equiv \frac{i J}{|V_{ud} V_{us}|} \Bigl[ \frac{2}{3} y_5 +2 y_6 \Bigr]$, with the Jarlskog invariant $J= (3.06^{+0.21}_{-0.20} )  \times 10^{-5}$ \cite{pdg,jarlskog}.
Using the partially conserved axial current formula, the pseudoscalar meson matrix elements are expressed as
\begin{eqnarray}
\langle \pi^0 | \bar d \gamma_5 d | 0 \rangle
&\approx &
- \frac{i}{f_\pi} \langle 0 | \bar q q | 0 \rangle
,
\label{eq:pi0vacuum}
\\
\langle \bar K^0 | \bar s \gamma_5 d | 0 \rangle
&\approx &
\frac{i}{f_K \sqrt{2} } \langle 0 | \bar q q + \bar s s | 0 \rangle
,
\label{eq:k0vacuum}
\\
\langle \eta | \bar d \gamma_5 d | 0 \rangle
&\approx &
\frac{i}{f_\eta \sqrt{3}} \langle 0 | \bar q q | 0 \rangle
,
\label{eq:etaddvacuum}
\\
\langle \eta | \bar s \gamma_5 s | 0 \rangle
&\approx &
- \frac{2i }{f_\eta \sqrt{3}} \langle 0 | \bar s s  | 0 \rangle
,
\label{eq:etassvacuum}
\end{eqnarray}
where the decay constants are given by $f_\pi = 93$ MeV, $f_K = 1.198 f_\pi $, and $f_\eta \approx f_{\eta_8} = 1.2 f_\pi $ \cite{pdg}.
The chiral condensates are given in Eq. (\ref{eq:chiralcondensate}).

For the baryon scalar matrix elements, we use those of Eqs. (\ref{eq:sdlamneu}), (\ref{eq:sdsigneu}), (\ref{eq:sdsigpro}), and the following nucleon matrix elements:
\begin{eqnarray}
\langle n | \bar d d | n \rangle
&\approx&
5.4
,
\\
\langle p | \bar d d | p \rangle
&\approx&
4.7
,
\end{eqnarray}
renormalized at the scale $\mu = 1$ GeV.
The above matrix elements were obtained by equating the phenomenological value of the pion-nucleon sigma term \cite{gasser,alarcon2,jlqcd1,dinter,durr2,yang2,qcdsf2,bhattacharya1,etm3}
\begin{eqnarray}
\frac{m_u + m_d}{2}
\langle N | \bar u u + \bar d d | N \rangle
&\approx&
45 \, {\rm MeV}
,
\end{eqnarray}
and the isovector nucleon scalar matrix element \cite{rpvbetadecay1,rpvedm,rpvbetadecay2,gonzales-alonso}
\begin{eqnarray}
\langle p | \bar u u - \bar d d | p \rangle
&=&
\frac{m_n^{(0)}-m_p^{(0)} }{m_d -m_u}
,
\end{eqnarray}
with $m_n^{(0)}$ and $m_p^{(0)}$ the proton and neutron masses without electromagnetic contribution \cite{thomas}.
The light quark masses $m_u = 2.9$ MeV and $m_d = 6.0$ MeV, at the renormalization scale $\mu =1$ GeV, were obtained through the two-loop level renormalization group evolution.

\end{document}